\documentclass[12pt]{article}

\usepackage{graphicx}

\begin{document}

\begin{center}
\vspace{24pt}
{\large \bf Second- and First-Order Phase Transitions in CDT}
\vspace{30pt}

{\sl J. Ambj\o rn}$\,^{a}$,
{\sl S. Jordan}$\,^{b}$,
{\sl J. Jurkiewicz}$\,^{c}$
and {\sl R. Loll}$\,^{b,d}$

\vspace{24pt}

{\footnotesize

$^a$~The Niels Bohr Institute, Copenhagen University\\
Blegdamsvej 17, DK-2100 Copenhagen \O , Denmark.\\
{ email: ambjorn@nbi.dk}\\

\vspace{10pt}

$^b$~Institute for Theoretical Physics, Utrecht University, \\
Leuvenlaan 4, NL-3584 CE Utrecht, The Netherlands.\\
{ email: s.jordan@uu.nl, r.loll@uu.nl}\\

\vspace{10pt}

$^c$~Institute of Physics, Jagellonian University,\\
Reymonta 4, PL 30-059 Krakow, Poland.\\
{ email: jurkiewicz@th.if.uj.edu.pl}

\vspace{10pt}

$^d$~Perimeter Institute for Theoretical Physics, \\
31 Caroline St. N., Waterloo, Ontario, Canada N2L 2Y5. \\
{email: rloll@perimeterinstitute.ca}

}

\vspace{48pt}

\end{center}

\begin{center}
{\bf Abstract}
\end{center}
Causal Dynamical Triangulations (CDT) is a proposal for a theory of 
quantum gravity, which implements a path-integral quantization of 
gravity as the continuum limit of a sum over piecewise flat 
spacetime geometries. 
We use Monte Carlo simulations to analyse the phase 
transition lines bordering the physically interesting de Sitter phase of
the four-dimensional CDT model.
Using a range of numerical criteria, we present strong
evidence that the so-called A-C transition is 
first order, while the 
B-C transition is second order. The presence of a second-order transition 
may be related to an ultraviolet fixed point of quantum gravity and thus provide 
the key to probing physics at and possibly beyond the Planck scale.

\newpage

\section{Introduction}\label{intro}

The relation between general relativity and quantum physics is still far 
from being understood, despite decades of research. Expanding the 
gravitational field around a fixed background geometry results in a 
perturbatively nonrenormalizable theory. The best one can do within
this conventional setting
is to view the quantum field theory as an effective
field theory up to a certain energy scale 
\cite{Donoghue:1995cz, Burgess:2003jk}. 
However, because of their extreme smallness, it remains unclear whether 
any quantum corrections one can
compute in the range where this effective framework is 
reliable can be related to observable quantities, 
even in principle \cite{Rothman:2006fp}.  

In order to define a theory which is UV-complete
one has to go beyond ordinary perturbative quantum field theory,
either by modifying or extending the notion of ``quantum fields" (as happens in
supergravity, string theory or noncommutative field theory, say), or by trying to 
define a quantum field theory of gravity {\it non}perturbatively 
(as, for example, in lattice or loop quantum gravity).
In this article we will be following a nonperturbative route to quantization
via the method of Causal Dynamical Triangulations (CDT), which falls into the
latter category. 

One explanation for why quantum gravity could be 
well-defined nonpertur\-ba\-tively,
despite its lack of perturbative renormalizability, is given by the 
so-called asymptotic safety scenario. Its general framework was
originally formulated by Weinberg \cite{Weinberg:1980gg}, and is based
on the hypothesis that the quantum field theory of four-dimensional
gravity possesses a non-Gaussian ultraviolet fixed point under the 
flow of the renormalization group. The existence of such a UV fixed 
point is suggested by a $(2\! +\!\epsilon)$-expansion in the dimensionality of
spacetime, starting from
the observation that two-dimensional ``gravity'' is formally a renormalizable
theory \cite{twopluseps}, an 
observation later corroborated in \cite{kawai}
(for a summary of these results, see \cite{nieder}). 
Further support, which does not rely on perturbing around $d=2$, 
comes from using functional renormalization group methods, first 
applied in the context of gravity in a seminal paper by Reuter
\cite{Reuter:1996cp}. Explicit numerical computations of the renormalization
group trajectories require a truncation of the infinite tower of higher-order
curvature 
terms in the gravitational Lagrangian. By now, numerous truncations have
been studied, which all point to the existence of a fixed 
point \cite{reuteretc} (see \cite{Reuter:2007rv} for a recent review).

Independent evidence for the asymptotic safety of gravity -- if it is
realized -- could come from 
a nonperturbative path integral 
quantization of gravity. In order to make the path integral well defined
one usually discretizes the space of field configurations, which for 
pure gravity is the space of all spacetime geometries. 
This can be conveniently achieved by representing spacetime geometries
in terms of piecewise flat, 
triangulated manifolds with distributional curvature
assignments, leading to the simplicial approaches 
to quantum gravity, such as quantum Regge calculus and dynamical 
triangulations (DT) (see \cite{Loll:1998aj} for a review).
In what follows we will study an improved, Lorentzian version of the latter, 
known as Causal Dynamical Triangulations (CDT) (see \cite{reviews}
for reviews).

A crucial and nontrivial requirement for all nonperturbative
approaches to quantum gravity is that they must reproduce
well-established classical behaviour of 
gravitational physics in the low-energy limit. 
In this respect CDT has had some remarkable successes. 
Numerical simulations of the model have demonstrated the 
dynamical emergence of a four-dimensional classical universe as 
the ground state of the system, even though no background was ever 
put in by hand \cite{Ambjorn:2004qm}. In addition, a more detailed 
analysis revealed that the effective action for the scale factor of the
universe can be matched to a minisuperspace action 
describing a quantum de Sitter spacetime, including 
linearized quantum fluctuations \cite{Ambjorn:2008wc}.
By contrast, the present piece of work will focus on the short-distance
behaviour of quantum gravity, defined through CDT. 

In the following section we will recap some essential ingredients of
the CDT approach, to set the stage for the subsequent
discussion of its phase structure. In Sec.\ \ref{section:phasediagram}
we recall some properties of the phase diagram and describe our
method of locating the phase transitions quantitatively. 
In Sec.\ \ref{sec:ptorder} we turn to the determination of the order
of two of the phase transitions, the transition line 
between phases A and C, and between
phases B and C. The fact that the B-C transition appears to be of second
order has been announced recently in \cite{phase1}.
Finally, our conclusions are presented in 
Sec.\ \ref{conclusion}.

\section{Causal Dynamical Triangulations}
\label{cdtsummary}

We begin by summarizing some important aspects of CDT, which are relevant 
for our subsequent discussion of its phase structure, and recommend consulting 
\cite{Ambjorn:2005qt,Ambjorn:2001cv,Ambjorn:2007jv} for more detailed information.
The primary idea 
behind CDT is to quantize gravity in the path-integral formalism, 
and is best illustrated in terms of the analogous 
quantization of the nonrelativistic particle. In order to properly define the 
path integral for the latter one can discretize time by introducing a unit time 
interval of length $a$, and consider only paths which consist of a contiguous 
sequence of linear segments, one for each time interval. Continuum quantum
physics is then recovered 
by taking the limit as $a\rightarrow 0$ of the regularized path integral over
the ensemble of piecewise linear paths. 

When generalizing from nonrelativistic particle physics to quantum gravity, 
the sum over piecewise linear (read: piecewise straight) particle paths 
becomes a sum over 
piecewise linear (read: piecewise flat) spacetime geometries. 
At this point one needs a concise definition of the (regularized) 
space of geometries to be summed over. 
The distinguishing feature of CDT, compared to other,
similar approaches, is that all geometries in its configuration space 
are equipped with a discrete foliation. 
Each leaf -- labeled by a discrete lattice time $t_n$ -- is 
a piecewise linear spatial hypersurface 
of a given, fixed topology $\mathcal{T}$. More precisely,
each spatial hypersurface is a simplicial manifold built out of equilateral 
flat tetrahedra 
with link length $a_s$, together forming a three-dimensional triangulation. 
Adjacent spatial hypersurfaces are then connected with the help of four-simplices, 
in such a way that the whole spacetime becomes a four-dimensional 
simplicial manifold of topology $\mathcal{T}\times[0,1]$. 
In practice we use $\mathcal{T}=S^3$ and furthermore impose 
periodic boundary conditions in the time direction, such that the spacetime 
topology becomes $S^3\times S^1$, but this should be regarded merely as a choice of
convenience.

The resulting four-dimensional triangulation forms a simplicial complex, 
consisting of vertices, links, triangles, tetrahedra and four-simplices. 
The Lorentzian nature of the metric properties of the triangulations implies
a more refined categorization of 
their elementary building blocks. There are two different 
types of links, depending on whether their endpoints lie on the same 
hypersurface (leading to \emph{spatial} links) or on two neighbouring 
hypersurfaces (resulting in \emph{timelike} links). In CDT quantum gravity
we allow these two types 
to have different absolute length and define the (squared) 
length of the timelike links to be $a_t^2=\alpha a_s^2$, with a relative
scaling parameter $\alpha<0$. Consequently, higher-dimensional
(sub-) simplices of a CDT geometry also come in different varieties,
depending on the space- or timelike character of their 
one-dimensional subsimplices.

The path integral now becomes a sum over all geometrically inequivalent 
triangulations of this type with a fixed number of time steps. 
As an action we use the Einstein-Hilbert action, which has a 
natural realization on piecewise linear geometries,
the so-called Regge action \cite{Regge:1961px}. The corresponding model in 
two dimensions can be solved analytically \cite{Ambjorn:1998xu}, 
but already the three-dimensional model
has only been solved partially and for restricted classes of 
triangulations \cite{3dmatrix,Benedetti:2007pp}. 
In four dimensions analytical methods are mostly unavailable and 
one must resort to Monte Carlo simulations to extract physical results. 
To do so one needs to convert the path integral into a 
statistical partition function by applying a Wick rotation. 

Because of the presence of a foliation and an associated global notion of time 
one can perform a Wick rotation 
at the level of the geometries by simply rotating 
$\alpha\rightarrow -\alpha$ in the lower-half complex plane. 
The Regge action changes accordingly and becomes the Euclidean Regge action
\begin{eqnarray}
S_E &=&\frac{1}{G}\int d^4 x\sqrt{g}(-R+2\Lambda) \nonumber \\
&\rightarrow& -(\kappa_0+6\Delta)N_0+\kappa_4 N_4+
\Delta (N_4+N_4^{(4,1)}), 
\label{reggeaction}
\end{eqnarray}
where $N_0$, $N_4$ and $N_4^{(4,1)}$ denote the numbers of vertices,
four-simplices and four-simplices of type (4,1) (with four vertices on one 
hypersurface and the fifth on a neighbouring hypersurface). 
The three couplings $\kappa_0$, $\kappa_4$ and $\Delta$ can be
expressed as functions of the bare gravitational coupling, 
the bare cosmological coupling and the asymmetry parameter $\alpha$ 
introduced above. By defining $\widetilde{\kappa_4}=\kappa_4+\Delta$ 
one obtains the version of the Euclidean Regge action which is ultimately 
implemented in the computer simulations, namely,
\begin{equation}
S_{Regge} = 
-\kappa_0 N_0 +\widetilde{\kappa_4} N_4 +\Delta (N_4^{(4,1)}-6 N_0),
\label{softwareaction}
\end{equation}
where we have grouped together the terms proportional to $\Delta$.
For later convenience we introduce the notation 
$\mathrm{conj}(\Delta)\! :=\! N_4^{(4,1)}\! - 6 N_0$
for the quantity conjugate to $\Delta$ in the action.

In this way the path integral of quantum gravity is turned into a statistical 
partition function. We will use the freedom to switch to a 
different ensemble, obtained by holding fixed $N_4$ -- measuring the
size (the total discrete spacetime volume) of the system -- instead of 
its conjugate $\kappa_4$.
We therefore treat $N_4$ as a finite-size scaling parameter which does 
not appear in the phase diagram of the putative continuum theory. The 
remaining two couplings $\kappa_0$ and $\Delta$ span the phase diagram
of the system, which we will explore in the next section.

\section{The phase diagram of CDT}
\label{section:phasediagram}

\begin{figure}[t]
\centerline{\scalebox{0.17}{\includegraphics{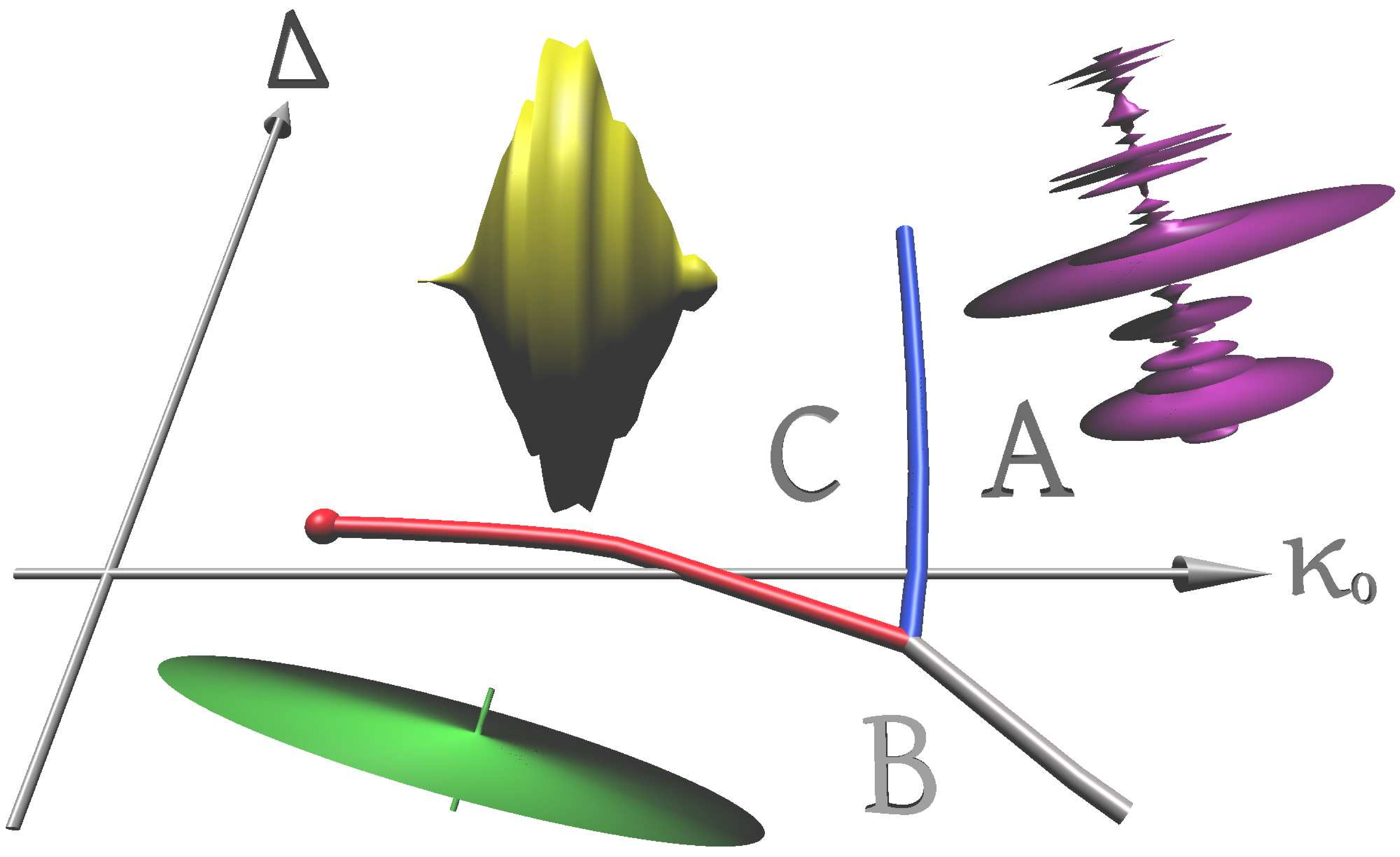}}}
\caption{Visualization of the phase diagram of CDT quantum gravity, 
based on actual measurements, illustrating the distinct
volume profiles characterizing the three observed phases A, B and C.}
\label{fig:pd_visualized}
\end{figure}

The first, qualitative description of the CDT phase diagram appeared  
in \cite{Ambjorn:2005qt}, followed more recently by a quantitative
and more detailed analysis in \cite{Ambjorn:2010hu}. 
The new visualization of the phase diagram shown in 
Fig.\ \ref{fig:pd_visualized} is based on phase transition data from the latter.
The phase diagram contains three phases, 
labelled A, B and C \cite{Ambjorn:2005qt}. They can be
distinguished by looking at a particular feature of the large-scale geometry
of their ground state, the spatial volume profile $N_3(t)$, which measures 
the three-volume in lattice units as a function of proper time $t$. 

Representative profiles have been visualized in 
Fig.\ \ref{fig:pd_visualized} by rotating the function $N_3(t)$ 
around the $t$-axis, thereby creating a body of revolution.
The average geo\-metry (``average" in the sense of expectation values)
found in phase C exhibits
the scaling behaviour of a genuinely four-dimensional universe, whose
average volume profile is consistent with a Euclidean 
de Sitter spacetime \cite{Ambjorn:2008wc,Ambjorn:2007jv}. 
The situation 
in the other phases is very different. The typical volume profile 
of a configuration in phase A consists of an essentially uncorrelated 
sequence of 
spatial slices, while the configurations in phase B are characterized 
by an almost vanishing time extension, in the sense that almost the
entire volume of the system is concentrated around a single spatial slice.

We should mention at this stage that several qualitative features of 
the CDT phase diagram bear a striking resemblance to those of a
similar diagram in
so-called Ho\v rava-Lifshitz gravity \cite{horava}, as has been pointed out elsewhere
\cite{HLphase,Ambjorn:2010hu}. Trying to understand whether this 
indicates a deeper relation
between the two quantum-gravitational frameworks provides an 
additional incentive for performing a more detailed quantitative analysis of
the phase diagram, as we are doing here.

\begin{figure}[t]
\centerline{\scalebox{1.3}{\includegraphics{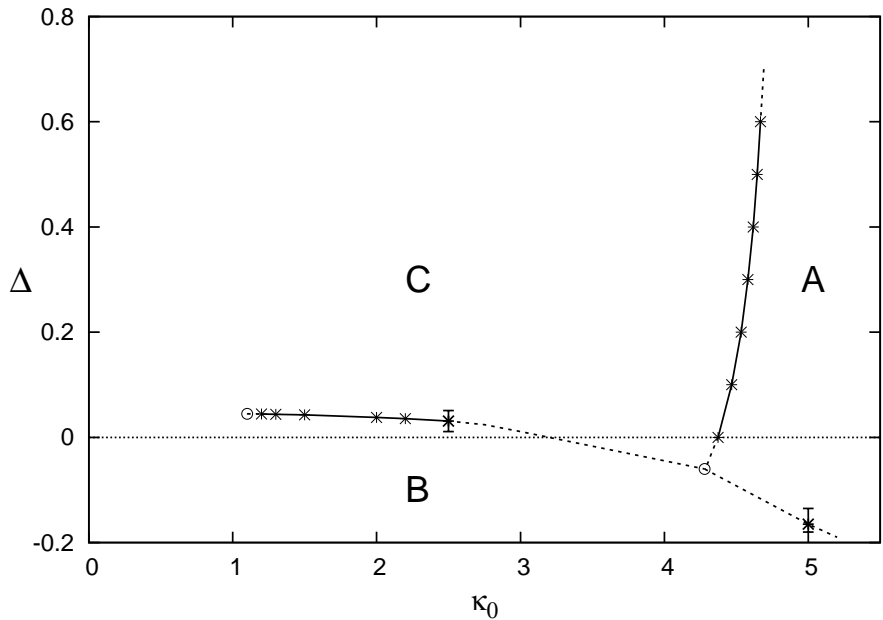}}}
\caption{Phase diagram of CDT quantum gravity \cite{Ambjorn:2010hu}. 
Crosses represent actual 
measurements, dashed lines are extrapolations.}
\label{fig:pd_plot}
\end{figure}

Fig.\ \ref{fig:pd_plot} shows the quantitative phase diagram, 
measured at system size $N_4=80.000$. The location of the 
phase transitions depends on the system size: when $N_4$ is increased, 
the A-C transition moves towards larger values of $\kappa_0$, 
while the B-C transition moves towards larger values of $\Delta$. 
Using finite-size scaling techniques it is possible to determine the location of 
transition points for infinite four-volume. 

Before embarking on a detailed analysis of the CDT phase transitions,
let us comment on some technical issues with regard to
measuring the location of a phase transition. The very 
notion of a phase transition point is of course ambiguous for finite systems, 
because strictly speaking phase transitions can only occur in the limit of infinite size.
For example, consider the susceptibility 
$\chi_{\cal O}=\left<{\cal O}^2\right>-\left<{\cal O}\right>^2$ associated with some observable $\cal O$. 
Plotting the function $\chi_{\cal O}(\kappa_0=\mathrm{const},\Delta)$ 
near the B-C transition, one will find that it has a maximum at some value $\Delta_c$. 
The location of this maximum can be used as a definition of the 
transition point, but one must keep in mind that its precise value will 
in general depend weakly on $\cal O$.

\begin{figure}[t]
\centerline{
\scalebox{1.0}{\includegraphics{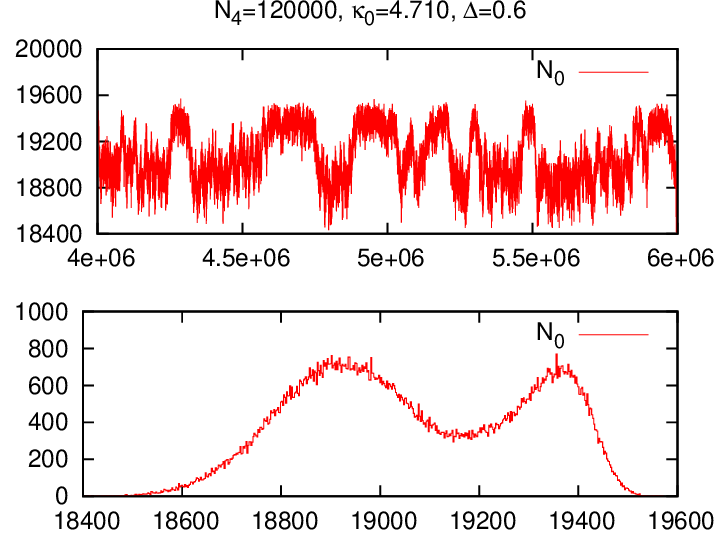}}
\scalebox{1.0}{\includegraphics{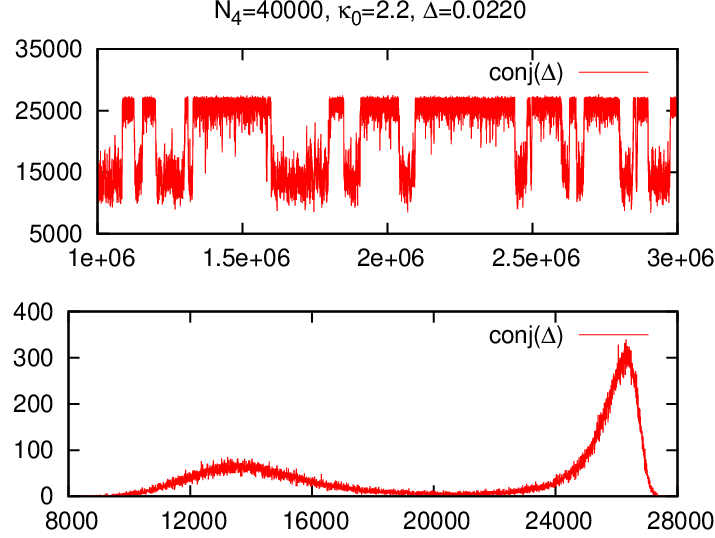}}
}
\caption{Monte Carlo time evolution (i) of $N_0$ at the A-C transition 
(top left) with associated histogram (bottom left), 
(ii) of $\mathrm{conj}(\Delta)$ 
at the B-C transition (top right) with associated histogram (bottom right).}
\label{fig:coex}
\end{figure}

Another definition of phase transition comes from studying histograms of suitable observables. Consider the Monte Carlo time evolution of the quantity $N_0$ near 
the A-C transition, 
as depicted in Fig.\ \ref{fig:coex} (top left). 
We observe that $N_0$ for some time fluctuates around one value, 
then makes a transition to another value, around which it fluctuates 
for a while, before flipping back to the 
first value again. This behaviour is characteristic of a phase transition, 
with the two different states corresponding to the two phases involved. 
The associated histogram of $N_0$ (Fig.\ \ref{fig:coex}, bottom left) shows a 
double-peak structure. Varying $\kappa_0$ while 
holding $\Delta$ fixed will change the relative heights of the two peaks. 
We can now fine-tune $\kappa_0$ so that both peaks have the same height. 
This gives us another definition of the phase transition point, although
often not the most practical one, since the fine-tuning can be  
very time-consuming. 
Also, there may be regions of coupling constant space or volume 
sizes where the flipping between the two states is not strong 
enough, in which case the histogram will not exhibit two peaks, but look more 
like a deformed Gaussian distribution.  

As we move towards $N_4\rightarrow\infty$, all of these definitions will
converge to one and the same phase transition point at infinite four-volume. 
However, what is important for our purposes is the observation that already 
for moderately large system sizes those definitions will produce values 
so close to each other as to be practically indistinguishable on a plot like
that of Fig.\ \ref{fig:pd_plot}. In order to determine the phase diagram it 
is therefore sufficient to fine-tune the parameters until a flipping of phases is 
observed. Not all observables are equally suited for this kind of analysis, 
because the flipping between phases can occur with different amplitudes for
different observables. A good choice is usually the quantity conjugate 
to the coupling that needs to be fine-tuned. The form of the action 
(\ref{softwareaction}) suggests using the quantity $N_0$ for the 
A-C transition, and $\mathrm{conj}(\Delta)=N_4^{(4,1)}\! -\! 6 N_0$  
for the B-C transition. The Monte Carlo time evolution and associated histogram 
for the latter are shown in Fig.\ \ref{fig:coex} on the right (top and bottom).

It is instructive to track how the phase transition signature changes as one 
moves along the phase transition lines, while holding the system size 
$N_4=80.000$ fixed. Along the A-C transition we have not observed any 
appreciable change, although we have not investigated the immediate
neighbourhood of the triple point 
where all three phases meet. The situation for the 
B-C transition is rather different. 
Let us start on the B-C line at the point $\kappa_0=2.2$ and move in both 
directions along the line. As we move to the left, the jump in 
$\mathrm{conj}(\Delta)$ associated with the phase flip 
decreases, and around $\kappa_0=1.0$ no observable signature of a 
phase transition remains. We conclude that the 
B-C transition line has an endpoint, which for $N_4=80.000$ is located 
around $\kappa=1.0$. Its precise location remains to be determined.

Conversely, when moving to the right from $\kappa_0=2.2$, we observe 
that the jump in $\mathrm{conj}(\Delta)$ grows rather quickly, until
at $\kappa_0=2.5$ we no longer observe the phase flipping, 
although the reason for this is very different from the situation 
encountered at the left endpoint. The increased size of the jump in 
$\mathrm{conj}(\Delta)$ is accompanied 
by a deepening of the vertical gap separating the two peaks of the histogram of
$\mathrm{conj}(\Delta)$. As it becomes deeper, the frequency of the 
phase flipping decreases significantly, until it becomes so small that 
no flipping is observed during the entire simulation. This indicates that
the simulation becomes stuck in a metastable state and does not 
reach thermal equilibrium. In Fig.\ \ref{fig:pd_plot} this region is marked
by a dashed line; here conventional methods are 
insufficient to measure the phase transition location with acceptable 
accuracy. We are currently investigating the use of multicanonical 
Monte Carlo simulations \cite{Berg:1998nj}, which at least in principle 
can offer a solution to this problem.

\section{How to measure the order of phase transitions}
\label{sec:ptorder}

Measuring the order of phase transitions requires some care, as
is illustrated by the history of dynamical triangulations. The quantum
gravity model based on four-dimensional {\it Euclidean} triangulations 
(which lack the foliation present in CDT) has a phase transition, which 
initially was determined to be of second order 
\cite{ambjornxx,Catterall:1994pg,Ambjorn:1995dj}. 
However, later studies with larger systems  
established instead that the phase transition is of first 
order \cite{Bialas:1996wu,deBakker:1996zx}.

Criteria to distinguish between first- and 
second-order transitions have been summarized conveniently
in \cite{MeyerOrtmanns:1996ea}, although care should be taken 
in extrapolating from conventional systems, simulated on static lattices, 
to dynamical triangulations, where the lattice itself is dynamical. 
Let us begin by considering the situation of infinite system size. In this
case first-order transitions are uniquely characterized by the existence of 
observables (first derivatives of the free energy) that are discontinuous 
across the transition. In addition, the size of the fluctuations relative 
to the average tends to zero in the infinite-volume limit, implying that the 
probability distribution of such an observable approaches the sum of two 
delta-function distributions. The displacement 
between the two peaks is precisely 
the difference between the two equilibrium values of the observable.

The probability distribution of an observable in the infinite-volume limit 
can be understood as arising through a limiting process of the 
corresponding distributions 
for finite volume. Their measured counterparts are histograms of the 
kind displayed in Fig.\ \ref{fig:coex}. It is sometimes stated in the literature that a 
double-peak structure of a histogram signals a first-order 
transition, but this is somewhat misleading and can even be wrong. 
Rather, the existence of a double-peak histogram 
allows one to confirm the first-order nature of a transition by 
considering a sequence of histograms 
for increasing system size. We are dealing with a first-order transition
whenever the double-peak structure becomes more pronounced
with increasing volume. Quantitatively, this means that the vertical
gap associated with the double peak increases. By ``vertical gap"
we mean the difference between the peak 
heights (assuming they are equal) and the height of the minimum in 
between them.
How the vertical gap changes with system size is the method of 
choice to confirm 
the first-order nature of a transition, provided one can 
simulate sufficiently large systems, which develop a double-peak 
structure. One could in principle try to use the same method to conclude 
that a transition is of second or higher order by measuring
the vertical gap as a function of inverse system size and extrapolating to 
the limit of infinite volume, where it should vanish. 
However, because it is often difficult to measure this gap with 
good accuracy, showing conclusively that it really goes to zero in the limit 
can be problematic. Fortunately, 
there are better criteria at hand -- involving the 
measurement of critical exponents -- which one can use to
establish that a given transition is of higher order.

One such exponent measures the shift of a transition point with system size. 
Recall first how this works for a conventional lattice system such as the 
Ising model with volume $V=L^d$, where $d$ is the system's dimension 
\cite{Newman:1999a}. Considering the temperature-driven phase 
transition of the Ising model and using the location of the maximum of 
the magnetic susceptibility to define a transition point $\beta^c(V)$,
one finds a power-law behaviour
\begin{equation}
|\beta^c(\infty)-\beta^c(V)|\propto V^{-1/\nu d}
\label{eq:ptshift}
\end{equation}
for sufficiently large system size.
The exponent $\nu$ governs the increase of the correlation length in 
a second-order transition as one moves towards the 
critical point $\beta^c(\infty)$ on an infinite lattice. 
For first-order transitions 
there is no correlation length and one expects the specific 
scaling \cite{MeyerOrtmanns:1996ea} 
\begin{equation}
|\beta^c(\infty)-\beta^c(V)|\propto V^{-1/\tilde{\nu}},~~~~\tilde{\nu}=1.
\label{eq:firstordershift}
\end{equation}
A sufficiently strong violation of $\tilde{\nu}=1$ therefore
signals the presence of a second-order transition.

To work with the criteria (\ref{eq:ptshift}) or (\ref{eq:firstordershift}) 
one needs some way of judging whether the system sizes under
consideration are large enough. 
Given $N$ data points ordered by system size, we can make $M$ fits where the 
$i$-th fit, $i\!\in\! \{1,\ldots ,M\} $, is made by using the restricted set
of data points with labels $i,\dots, N$. 
If the corresponding sequence of exponents is drifting, 
it can be an indication that the system sizes are not 
large enough. Obviously, this method is only useful when one 
has sufficiently many data points. 

Another quantity of interest is the so-called Binder cumulant associated
with an observable $\cal O$, which may be 
defined as \cite{MeyerOrtmanns:1996ea}
\begin{equation}
\label{bind}
B_{\cal O}=\frac{1}{3}\left(1-\frac{\left<{\cal O}^4\right>}
{\left<{\cal O}^2\right>^2}\right)
= -\frac{1}{3} \;
\frac{\langle ({\cal O}^2)^2\rangle -\langle {\cal O}^2\rangle^2}
{\langle {\cal O}^2\rangle^2},
\end{equation}
and is always nonpositive. 
Considering $B_{\cal O}$ as a function of the couplings, 
its local minima are at transition points, where the fluctuations
are largest. We can measure these 
minima for different system sizes and by extrapolation determine 
$B_{\cal O}^{\min}(1/N_4\! =\! 0)$. This quantity is zero 
when the probability distribution of $\cal O$ 
approaches a delta function around
an expectation value $\langle {\cal O} \rangle$ in the 
infinite-volume limit, as is expected
at a second-order transition. On the other hand, many aspects 
of first-order transitions are described well by approximating the 
histogram of the observable with a superposition of two  
distributions centred at the expectation values 
$\langle {\cal O} \rangle\! =\! {\cal O}_1$
and $\langle {\cal O} \rangle\! =\! {\cal O}_2$ in the two phases. 
Again, if these distributions
approach delta functions in the infinite-volume limit, the minimum
of $B_{\cal O}$ will be given by
\begin{equation}\label{BO}
B_{\cal O}^{\min}(1/N_4=0) = -\frac{({\cal O}_1^2-{\cal O}_2^2)^2}
{12\, {\cal O}_1^2 {\cal O}_2^2},
\end{equation}
which is obtained for a value of the coupling constants where the relative
strength of the delta functions at ${\cal O}_1$ and ${\cal O}_2$ is given by 
${\cal O}_2^2/({\cal O}_1^2+{\cal O}_2^2)$ and 
${\cal O}_1^2/({\cal O}_1^2+{\cal O}_2^2)$ respectively. 
We conclude that a sufficiently strong deviation of $B_{\cal O}^{\min}$ 
from zero signals a first-order phase transition. 
Using the value of the Binder cumulant to make the case for a second-order 
transition is more difficult, 
because a weak first-order transition may show a convergence to a 
value close to zero. 

There are other methods, involving the measurement of other critical 
exponents, which can help to determine the order of a phase 
transition, but their translation to CDT quantum gravity leads to ambiguities.
We will therefore proceed by applying the methods described above to the 
A-C and B-C transitions found in the CDT system. The third transition (A-B) is 
currently of minor interest, since it bounds two phases
that most likely have no relevance for continuum physics.

\subsection{The order of the A-C transition}

\begin{figure}[t]
\centerline{\scalebox{0.8}{\rotatebox{-90}{\includegraphics{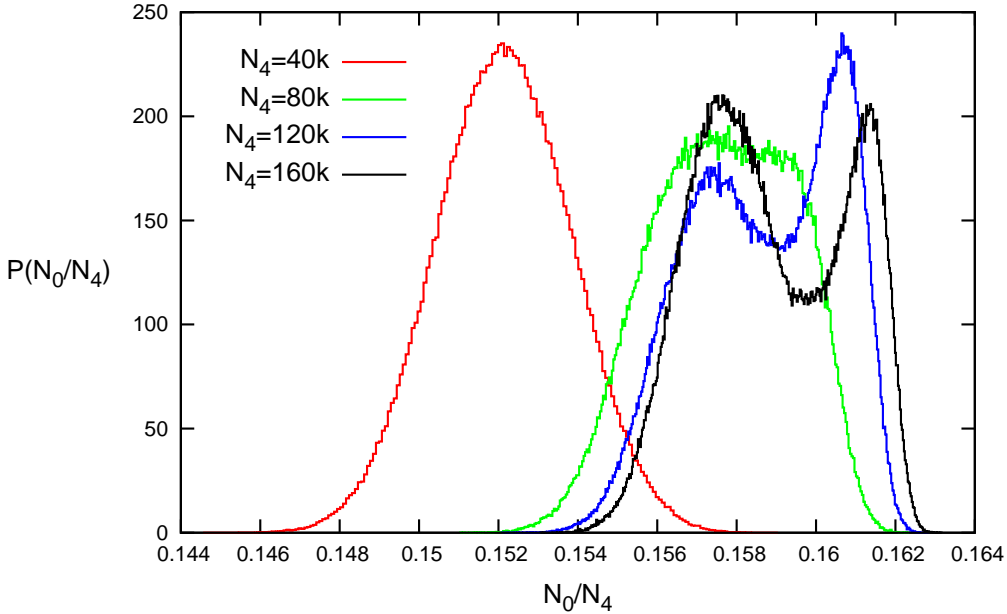}}}} 
\caption{Histograms of $N_0$ at the A-C transition 
for four different system sizes. 
The histograms have been normalized to obtain probability distributions $P$.}
\label{fig:n0histnorm}
\end{figure}

The simulations for the analysis of the A-C transition were run at $\Delta=0.6$, 
with system sizes ranging from $40k$ to $150k$.
The data presented in the following are the results of simulations 
where the number of sweeps was approximately $5\cdot 10^6$, with
one sweep representing one million attempted Monte Carlo moves.  

Our simulations suffer from a slow convergence of 
observables, which means that the standard error algorithms produce 
error values which significantly underestimate the true uncertainties 
of the measurements. In all our measurements we have attempted to 
obtain more realistic error values by systematically
studying the convergence of observables. 
Given a set of data samples at Monte Carlo times $\tau_1,\ldots ,\tau_n$, 
we can define a time-dependent average $\left<O\right>(\bar{\tau})$ by 
including only those data samples for which $\tau_1\leq \tau\leq \bar{\tau}$. 
We then plot this quantity as a function of $\bar{\tau}$ and try to extract 
reasonable error values. This method introduces a degree of subjectivity,
but in our opinion produces more realistic error estimates.

We first perform a histogram analysis. Fig.\ \ref{fig:n0histnorm} shows the 
histograms for the observable $N_0$ (more precisely, the rescaled
quantity $N_0/N_4$) at four different system sizes. 
The histogram at $N_4=40k$ assumes an approximate Gaussian shape, 
which gets distorted at $N_4=80k$, with a double peak starting to emerge
at $N_4=100k$, which becomes more pronounced at $N_4=120k$. 
For the two larger volumes, the vertical gap associated with the double-peak
clearly increases, while
the mutual distance of the peaks stays approximately the same.  
This behaviour is a clear signature of a first-order transition. 
The emergence of the double-peak shape seems to occur somewhere 
between $N_4\! =\! 80k$ and $N_4\! =\! 100k$. This may be compared  
with the analysis performed in the context of four-dimensional
Euclidean dynamical triangulations, 
where the double peak emerged between $N_4\! =\! 16k$ and $N_4\! =\! 32k$ 
\cite{Bialas:1996wu}. The more rigid structure imposed on 
the triangulations by the causality conditions of CDT may be responsible
for the shift of the appearance of the first-order signal
to somewhat larger volumes. 

Let us try to find additional evidence for the first-order nature of 
the A-C transition by first
measuring the shift exponent $\tilde{\nu}$ 
defined in eq.\ (\ref{eq:firstordershift}). 
For sufficiently large system sizes we expect a power-law behaviour 
\begin{equation}
\kappa_0^c(N_4)=\kappa_0^c(\infty)-C N_4^{-1/\tilde{\nu}},
\label{eq:ptshiftAC}
\end{equation}
where $\kappa_0^c(N_4)$ denotes the transition point at system size $N_4$ 
and $C$ is a proportionality factor. 

We have already mentioned that the notion of a transition point is 
ambiguous for finite system sizes. For our present purposes, we 
define $\kappa_0^c(N_4)$ as the location of the maximum of the susceptibility 
$\chi_{N_0}=\left<N_0^2\right>-\left<N_0\right>^2$.
To measure this maximum we use an 
extrapolation method due to Ferrenberg and Swendsen 
\cite{Ferrenberg:1988yz}, also known as histogram 
method \cite{Newman:1999a}.

\begin{figure}[t]
\centerline{\scalebox{0.55}{\includegraphics{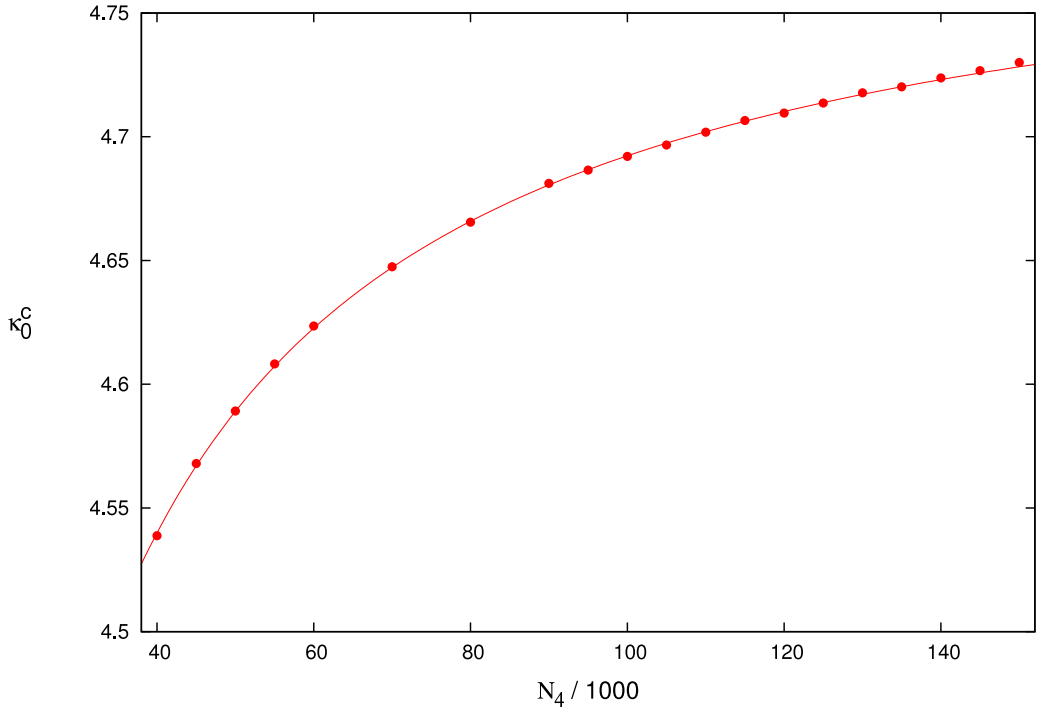}}}
\caption{Measured A-C transition points $\kappa_0^c(N_4)$ at 
$\Delta=0.6$ for different system sizes $N_4$, together with 
a best fit to the inverse-volume expansion eq.\ (\ref{firstorderfit}).}
\label{firstorder}
\end{figure}
The best fit using all data points gives $\tilde{\nu}=\! 1.11(2)$,
with a reduced chi-squared of $\chi^2\! =\! 2.1$. 
Unfortunately, the three-parameter 
fit (\ref{eq:ptshiftAC}) is not very stable when removing the points with the 
lowest values of $N_4$. Taking this into account,
it becomes clear that the error bar of 0.02 significantly
underestimates the error
associated with the determination of $\tilde{\nu}$ from the 
data. It indicates the presence of subleading terms which for the
volume range studied interfere with the pure power-law behaviour
of (\ref{eq:ptshiftAC}). Consequently, the strongest conclusion we
can draw from
using this method is that the measured $\tilde\nu\! =\! 1.11$ 
is compatible with the value  
$\tilde{\nu}\! =\! 1$ characteristic of a first-order transition.

As an additional cross-check 
we have made another three-parameter fit to the generic functional
form of a large-$N_4$ expansion of $\kappa_0^c(N_4)$, valid at 
a first-order transition \cite{MeyerOrtmanns:1996ea}, namely,
\begin{equation}\label{firstorderfit}
\kappa_0^c(N_4) = \kappa_0^c(\infty) -C/N_4 -D/N_4^2 + O(1/N_4^3).
\end{equation}
Fig.\ \ref{firstorder} shows the measured
data points and the best fit, based on the expansion (\ref{firstorderfit}). 
The errors associated with individual data points are too small to be 
visualized in a sensible way. 
The quality of this fit is comparable with that 
obtained from fitting to the power law (\ref{eq:ptshiftAC}).
The contribution from the $1/N_4^2$-term is subdominant 
to that from the $1/N_4$-term, as should be.
To summarize, we can say with confidence that 
the shift data are fully compatible with a first-order transition. 
However, it is also clear that if one wanted to go beyond this statement
and nail down the value of the critical exponent $\tilde{\nu}$ with better
precision, one would need to consider larger system sizes.

\begin{figure}[t]
\centerline{\scalebox{1.3}{\includegraphics{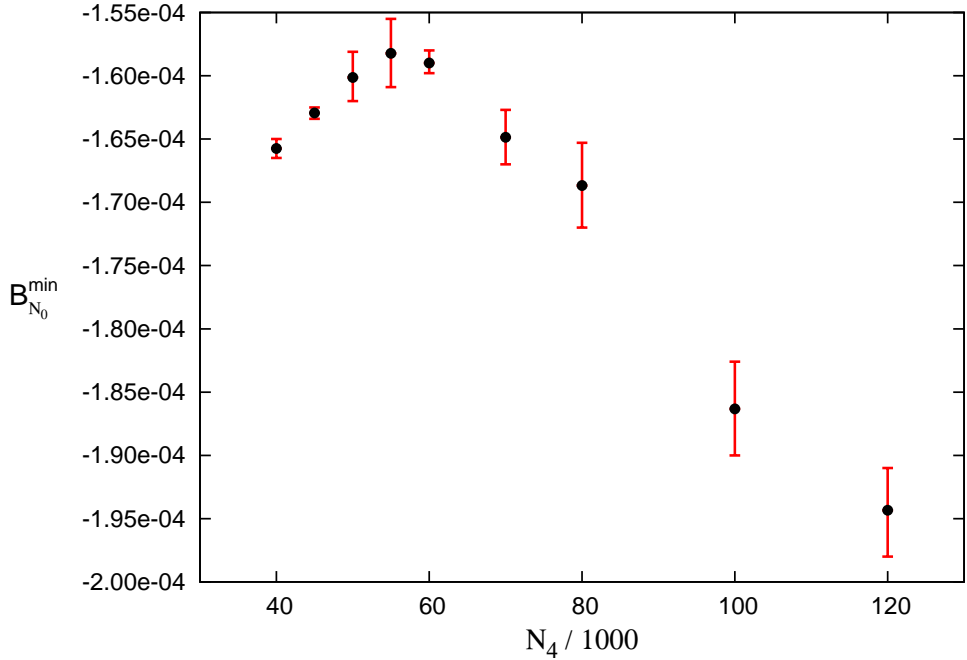}}}
\caption{Dependence of the minimum $B_{N_0}^{\min}$ of the 
Binder cumulant $B_{N_0}$ on the system size $N_4$, at the A-C transition and
for $\Delta \! =\! 0.6$.}
\label{fig:bincumAC}
\end{figure}

Our next and final step in analyzing the A-C transition 
will be to consider the Binder cumulant for $N_0$,
\begin{equation}
B_{N_0}=\frac{1}{3}
\left(1-\frac{\left<N_0^4\right>}{\left<N_0^2\right>^2}\right),
\end{equation}
which has a local minimum $B_{N_0}^{\min}$ at the A-C phase 
transition.
Like in the case of the shift exponent $\tilde\nu$, we use the histogram 
method to determine this minimum. In Fig.\ \ref{fig:bincumAC} we have plotted 
$B_{N_0}^{\min}$ as a function of system size. 
The error bars are much larger than for the measurement of the 
shift exponent, where we measured the \emph{location} of the 
susceptibility maximum, and not the maximum value itself.
The plot shows clearly that $B_{N_0}^{\min}$ moves away from zero as one 
goes to large system sizes, reconfirming the first-order nature of the 
transition. 

By contrast, when going to smaller volumes one finds a different behaviour,
where $B_{N_0}^{\min}$ increases before reaching a local maximum. 
Evidently, for these small volumes the system lies outside the scaling 
region, where quantities like $B_{N_0}^{\min}$ 
are expected to behave according to a power law. 
Looking at the plot, a rough estimate of the onset of the scaling region
seems to be around or above $N_4=60k$. Comparing with 
Fig.\ \ref{fig:n0histnorm}, this may be correlated with the observed
emergence of the double-peak structure in the histograms.

\subsection{The order of the B-C transition}

\begin{figure}[t]
\centerline{\scalebox{0.8}{\includegraphics{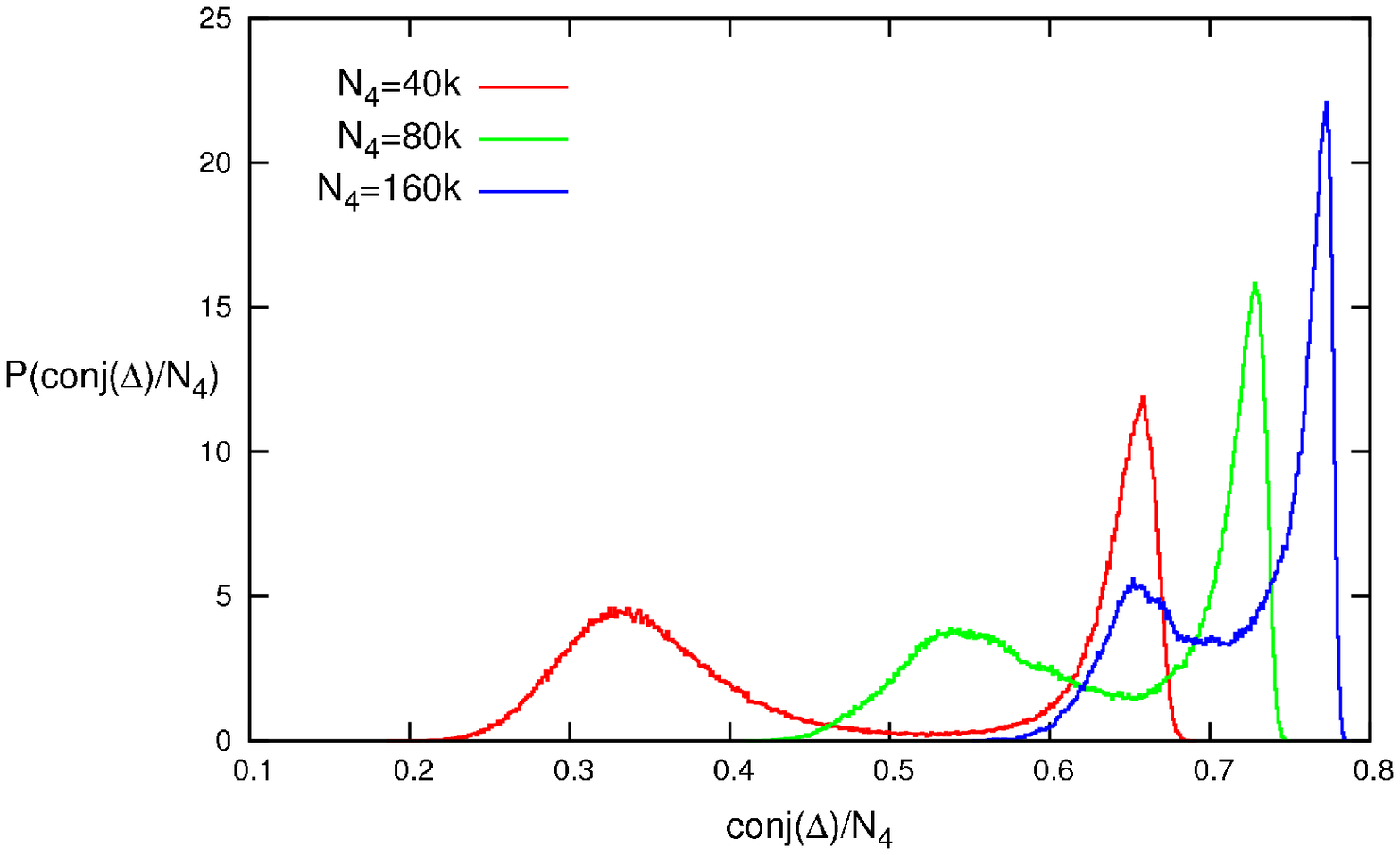}}}
\caption{Histograms of $\mathrm{conj}(\Delta)$ at the B-C transition 
for three different system sizes. 
The histograms have been normalized to obtain probability distributions $P$.}
\label{fig:histBC}
\end{figure}

Continuing our investigation of the order of phase transitions in
CDT quantum gravity, we now turn to the B-C transition.
In this case, we have fixed the inverse gravitational coupling to 
$\kappa_0\! =\! 2.2$ and analysed the system at
sizes between $40k$ and $160k$. The number of 
sweeps used was approximately $2.5\cdot 10^6$, with
one sweep again corresponding to one million attempted Monte Carlo moves.  
Fig.\ \ref{fig:histBC} shows the histograms of the quantity
$\mathrm{conj}(\Delta)\!=\! N_4^{(4,1)}\! - 6 N_0$ for three different system sizes. 
The situation differs from that of the A-C transition in that
we observe a double-peak structure for all four-volumes. The peaks
are most pronounced for smaller volume and appear to be merging 
when the volume is increased. 
The plot shows that their mutual distance decreases, roughly like
$1/N_4$. This is a first indication
that the B-C transition may not be a first-order transition.\footnote{Since 
we have not performed a 
time-consuming fine-tuning of $\Delta$ to obtain peaks of equal height,
we have not been able to extract a reliable estimate of their
vertical gap (relative height of peaks w.r.t. minimum in between peaks, as
defined earlier in Sec.\ \ref{sec:ptorder}).}

\begin{figure}[t]
\centerline{\scalebox{1.3}{\includegraphics{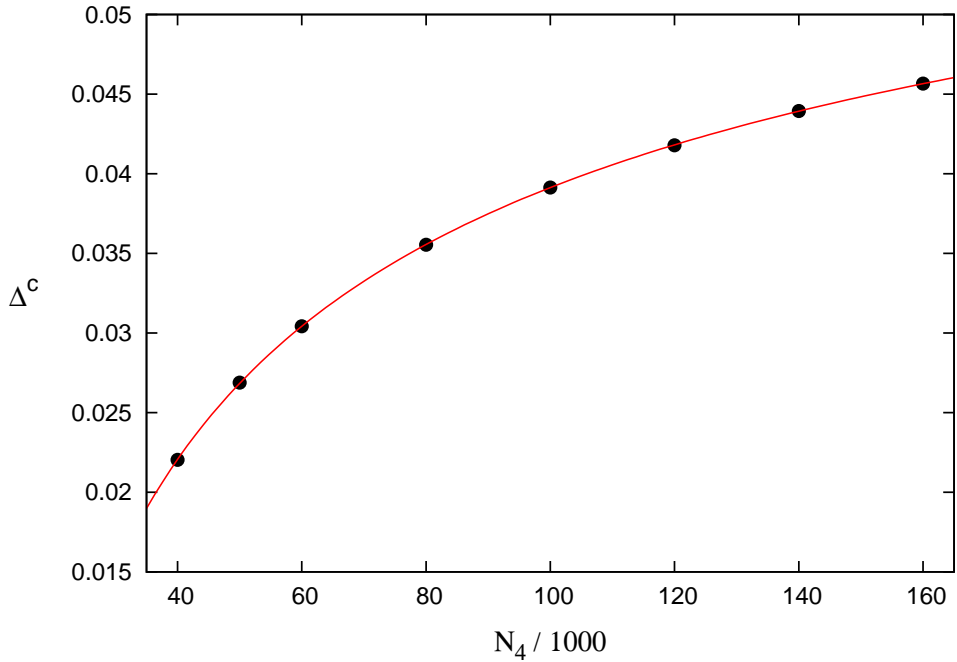}}}
\caption{Measured B-C transition points $\Delta^c(N_4)$ at 
$\kappa_0=2.2$ for different system sizes $N_4$, with a best fit   
to eq.\ (\ref{eq:ptshiftBC}) to determine the shift exponent $\tilde{\nu}$.}
\label{fig:shiftexpBC}
\end{figure}

We proceed by measuring the shift exponent for $\Delta$. 
Analogous to what we did for the A-C transition, we will use the formula
\begin{equation}
\Delta^c(N_4)=\Delta^c(\infty)-C N_4^{-1/\tilde{\nu}},
\label{eq:ptshiftBC}
\end{equation}
where $\Delta^c$ is defined as the location of the maximum of the 
susceptibility 
$\chi_{\mathrm{conj}(\Delta)}=
\left<\mathrm{conj}(\Delta)^2\right>\! -\!\left<\mathrm{conj}(\Delta)\right>^2$. 
Data points and best fit are displayed in Fig.\ \ref{fig:shiftexpBC}. 
We have not included any error bars 
because they turned out to be too small. The best fit through all 
data points yields $\tilde{\nu}=2.39(3)$. To judge whether 
our range of system sizes lies inside the scaling region we have again 
performed a sequence of fits by successively removing the data points 
with the lowest four-volume. The corresponding values for $\tilde\nu$
are $2.39(3)$, $2.51(3)$, $2.49(3)$ and $2.51(5)$,  
where the last fit was done with all but five data 
points removed. Again the error bars are based on making cuts in the 
sampled data as described above when discussing the A-C transition. 
The sequence suggests  
that the data point with the lowest four-volume lies 
outside the scaling region. Removing it from the fit we get
\begin{equation}
\tilde{\nu}=2.51(3).
\end{equation}
This result makes a strong case for a second-order transition, 
since the prediction $\tilde{\nu}=1$ for a first-order transition is 
clearly violated. 

\begin{figure}[t]
\centerline{\scalebox{1.3}{\includegraphics{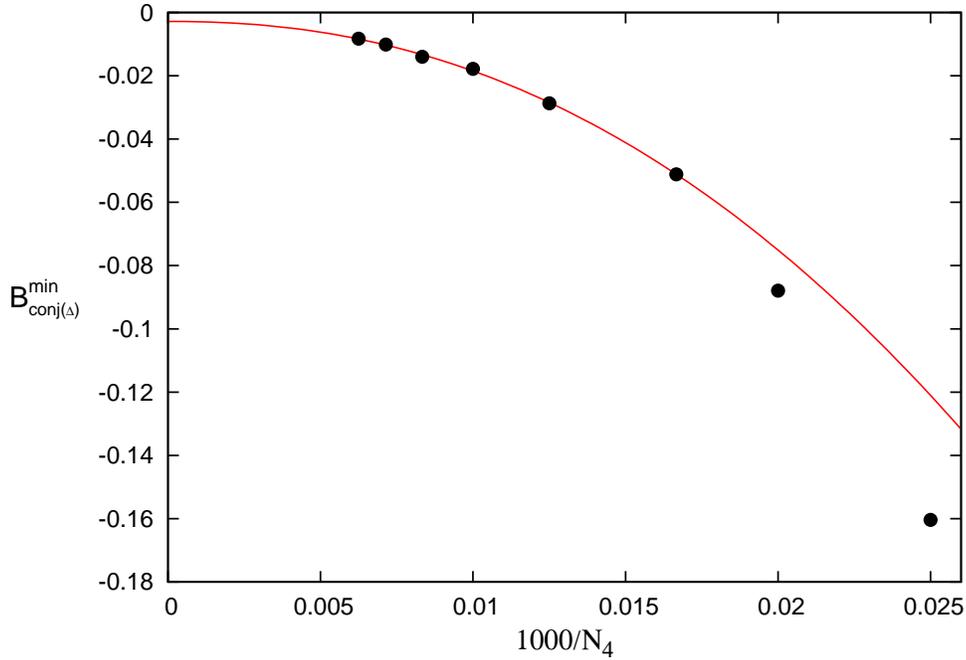}}}
\caption{Dependence of the minimum $B_{\mathrm{conj}(\Delta)}^{\min}$
of the Binder cumulant 
$B_{\mathrm{conj}(\Delta)}$ on the (inverse) system size $N_4^{-1}$,
at the B-C transition and for $\kappa_0=2.2$. 
(Fit excludes the two points on the right.)}
\label{fig:bincumBC}
\end{figure}

Let us finally consider how the minimum of the Binder cumulant $B_{\cal O}$
of definition (\ref{bind}), with ${\cal O}={\mathrm{conj}(\Delta)}$,
depends on the system size. In Fig.\ \ref{fig:bincumBC} we
show $B_{\mathrm{conj}(\Delta)}^{\min}$ 
as a function of \emph{inverse} system size. The errors turned out to be 
much smaller than for the corresponding measurements 
at the A-C transition, 
with error magnitude approximately equal to the radius of the dots in the plot. 
Inside the scaling region the minimum of the Binder cumulant is expected 
to behave like a power-law function. To determine whether our data 
points lie inside this scaling region, we have again performed a sequence of fits 
by successively removing the data points 
with the lowest four-volume. We have found that the data points at $N_4=40k$ and 
$N_4=50k$ both lie outside the scaling region. 
The curve displayed in Fig.\ \ref{fig:bincumBC} corresponds to the fit 
with those two points removed. 

\begin{table}[t]
\begin{center}
\begin{tabular}{|c|r|}
\hline
Observable $\cal O$ & $B_{\cal O}^{\min}(N_4\rightarrow\infty)$ \\
\hline
\hline
$\mathrm{conj}(\Delta)$ & $-0.003(4)$ \\
\hline
$N_4^{(4,1)}$ & $-0.001(3)$ \\
\hline
$N_2$ & $-0.000\, 000\, 1(3)$ \\
\hline
$N_1$ & $-0.000\, 003(7)$ \\
\hline
$N_0$ & $0.000\, 0(3)$ \\
\hline
\end{tabular}
\end{center}
\caption{Measurements of $B_{\cal O}^{\min}(N_4\rightarrow\infty)$ 
for various observables $\cal O$, where $N_k$ denotes the number of 
$k$-dimensional (sub-)simplices in the triangulation.}
\label{tab:bincumBC}
\end{table}

Table \ref{tab:bincumBC} collects the results of measuring 
$B_{\cal O}^{\min}(N_4\rightarrow\infty)$ for five different 
observables $\cal O$. All of these observables exhibit a phase
flipping at the transition, with associated double-peak histograms.
The interpolation to $N_4\! \to\! \infty$ is made by assuming 
that the quantities $B_{\cal O}^{\min}(N_4)$ for sufficiently large
volumes have a scaling behaviour like in eqs.\  
(\ref{eq:ptshiftAC}) and (\ref{eq:ptshiftBC}), namely,
\begin{equation}
\label{bogen}
B_{\cal O}^{\min}(N_4) = B_{\cal O}^{\min}(N_4\rightarrow\infty)-
C({\cal O})/N_4^{\tilde{\rho}({\cal O})},
\end{equation}
with $\cal O$-dependent exponents $\tilde{\rho}({\cal O})$ and coefficients
$C({\cal O})$. 
If we had to rely on the Binder cumulant measurements alone to determine
the order of the B-C transition, we would encounter the 
already mentioned problem of how to
distinguish between a second-order transition and a first-order transition
where $B_{\cal O}^{\min}$ converges to a value close to zero.
In the case at hand, in addition to the evidence already presented in
favour of a second-order transition,
the size and quality of the measurement
errors for the minimum of the Binder cumulants
are such that we can conclude with confidence that the results 
displayed in Table \ref{tab:bincumBC} are certainly {\it consistent}
with a limiting value of zero for $N_4 \to \infty$. This would be true even 
if the errors overestimated the true uncertainties by a factor of two, say.

\section{Discussion and outlook}
\label{conclusion}

We set out with the goal to determine the order of the two physically
relevant phase transitions in CDT quantum gravity. 
For conventional systems on static lattices, describing physics on a fixed
background geometry,
the methods for doing so are well established. 
It is not clear a priori whether all of them
are applicable for systems based on dynamical lattices, such as CDT,
which reflect the dynamical character of the quantum-gravitational,
geometric degrees of freedom they aim to describe.
Evidence that standard methods can be adapted rather straightforwardly
to systems of dynamical geometry comes from Euclidean dynamically
triangulated (DT) systems. This has been demonstrated for 
two-dimensional Euclidean quantum gravity (coupled to matter), where 
extensive computer simulations \cite{2dcomputer} are in
agreement with Liouville quantum gravity, which can be solved analytically.
It is also true in four-dimensional DT. Although this model does not appear
to describe four-dimensional quantum gravity, it {\it does} exhibit a
well-defined scaling behaviour in the so-called branched-polymer phase 
and finite-size scaling works very well \cite{Ambjorn:1995dj}. 

However, since causal dynamical triangulations differ from their purely
Euclidean counterparts both in their set-up
and (hopefully) the continuum physics they describe, we decided 
to not take the validity of these methods for granted. Rather, we used  
several independent criteria for determining the transition orders, 
and checked whether the results were mutually consistent.

Our analysis consisted of three parts:
the histogram analysis, the measurement of the shift exponent 
and the analysis of Binder cumulants. The histogram analysis
was sufficient to confirm the first-order nature of the A-C transition. 
It probes the definition of a first-order transition in a direct way and 
is therefore the method of choice to confirm the first-order nature of a 
transition. Measurements of the Binder cumulants also
pointed rather unambiguously to a first-order transition. The 
absolute values of the minima were increasing for increasing volume $N_4$,
contrary to what one would expect for a second-order transition.
Measurements of the shift exponent, which governs the dependence of 
transition points on the system size, were less conclusive. The results
were compatible with $\tilde{\nu}=1$, valid at a first-order transition.
However, since the stability of the curve fits under successive removal of data 
points with small volume was not particularly good, we did not succeed in 
{\it determining} the exponent $\tilde{\nu}$ convincingly
from the data, but only in establishing its consistency with the first-order
value.

The corresponding analysis of the B-C transition gave a very different
picture. The histogram analysis showed a double-peak structure,
but without the characteristic features of a first-order 
transition. The distance between the two peaks diminished with 
increasing $N_4$, indicating a higher-order transition.
This was corroborated by measuring the shift exponent, whose 
value $\tilde{\nu}=2.51(3)$ entails a strong violation of the 
prediction $\tilde{\nu}=1$ for a first-order transition. 
In addition, the results of the Binder cumulant 
analysis were also clearly consistent with the 
presence of a second-order B-C transition, as announced previously
in \cite{phase1}.

In summary, we have found strong evidence that in CDT quantum
gravity the A-C transition line 
is first order, while the B-C transition line is second order. 
The latter result is both remarkable and attractive. 
It opens the door to studying critical phenomena in CDT and
to defining a continuum limit for vanishing lattice spacing (UV cutoff).    
A discussion of the effective gravitational coupling constant, and its
dependence on the bare couplings of the model -- inside
phase C -- was initiated in \cite{Ambjorn:2008wc}.
In view of the results presented here, a future task will be to extend
this analysis by studying the flow of this coupling constant when 
approaching the second-order B-C phase
transition line. Apart from the intrinsic interest in understanding the
behaviour of quantum-gravitational observables in this limit, which
appears to be associated with probing (sub-)Planckian physics, 
this may enable one to make a more explicit connection 
with the asymptotic safety scenario, as outlined recently in \cite{physrep},
as well as anisotropic gravity models of Ho\v rava-Lifshitz type.

To make this connection more concrete, let us consider a potential
application of our analysis, based on the presence of a second-order transition.
As shown in \cite{Ambjorn:2004qm,Ambjorn:2005qt}, 
well inside phase C the time extent $T$ of the 
universe scales like $N_4^{1/4}$,
and the three-volume $V_3(t)$ of a generic spatial slice at time $t$ like 
$N_4^{3/4}$. By contrast, in phase $B$ this time extent 
essentially vanishes, since all three-volume is located at a single spatial slice. 
Approaching
the B-C transition line from inside phase C, we observe that $T$, measured in 
lattice time steps, decreases,
at least in the region $\kappa_0 
< 2.4$ where detailed measurements are available.
Its precise behaviour still needs to be determined, but it is 
natural to conjecture a behaviour of the form
\begin{equation}
T^c(N_4) \propto (\Delta^c(\infty)-\Delta^c(N_4))^\theta N_4^{1/4}
\label{scaling1}
\end{equation}   
at the pseudo-critical point $\Delta^c(N_4)$. Combining this with
the scaling (\ref{eq:ptshiftBC}) of the coupling $\Delta$, we obtain
\begin{equation}
T^c(N_4 ) = N_4^{1/4-\theta/\tilde{\nu}},
\end{equation}
implying that the time extent $T$ scales anomalously. 
Interestingly, such an 
anomalous scaling of time relative to space in the ultraviolet 
is also a key feature of Ho\v rava-Lifshitz gravities
\cite{horava}. Following this line of reasoning further, 
one could imagine
that both $\tilde{\nu}$ and $\theta$ depended on $\kappa_0$, in this 
way potentially allowing for both anisotropic and isotropic
UV completions, depending on where the B-C phase transition line is
approached.

\paragraph {Acknowledgements.} JA thanks the ITP and the Department of 
Physics and Astronomy of Utrecht University, as well as the Perimeter Institute 
for hospitality and financial support. He also acknowledges financial support
of the Danish Research Council under the grant "Quantum gravity and the role 
of black holes". JJ acknowledges partial support through the Polish Ministry 
of Science grants N N202 229137 (2009-2012) and 182/N-QGG/2008/0. 
SJ would like to thank Prof. G.T. Barkema 
for fruitful discussions and valuable advice concerning 
the numerical aspects of this work. RL acknowledges support by the Netherlands
Organisation for Scientific Research (NWO) under their VICI program.
The contributions by SJ and RL are part of the research 
programme of the Foundation for Fundamental Research 
on Matter (FOM), financially supported by NWO. This research was 
supported by Perimeter Institute for Theoretical Physics. Research at 
Perimeter Institute is supported by the Government of Canada through Industry
Canada and by the Province of Ontario through the Ministry of Research and
Innovation.

%\bibliographystyle{unsrt}
%\bibliography{pt.bib}

%\end{document}

%--------------------------------------------------------------------

\end{document}